\documentclass[twocolumn,preprintnumbers,amsmath,amssymb,superscriptaddress]{revtex4}
\usepackage{graphicx}
\usepackage{dcolumn}
\usepackage{bm}
\usepackage{float}
\usepackage{afterpage}

\newcommand{\comment}[1]{}

\usepackage{color}

\begin{document}

\title{Ultracold Heteronuclear Mixture of Ground and Excited State Atoms}

\author{Alexander Khramov}

\author{Anders Hansen}

\author{William Dowd}

\author{Richard J. Roy}
\affiliation{Department of Physics, University of Washington, Seattle WA 98195, USA}

\author{Constantinos Makrides}

\affiliation{Department of Physics, Temple University, Philadelphia PA 19122, USA}

\author{Alexander Petrov}

\affiliation{Department of Physics, Temple University, Philadelphia PA 19122, USA}
\affiliation{St. Petersburg Nuclear Physics Institute,
Gatchina, 188300; Division of Quantum Mechanics, St. Petersburg State University, 198904, Russia}

\author{Svetlana Kotochigova}

\affiliation{Department of Physics, Temple University, Philadelphia PA 19122, USA}

\author{Subhadeep Gupta}

\affiliation{Department of Physics, University of Washington, Seattle WA 98195, USA}

\date{\today}

\begin{abstract}

We report on the realization of an ultracold mixture of lithium atoms in the ground state and ytterbium atoms in an excited metastable (${^3P_2}$) state. Such a mixture can support broad magnetic Feshbach resonances which may be utilized for the production of ultracold molecules with an electronic spin degree of freedom, as well as novel Efimov trimers. We investigate the interaction properties of the mixture in the presence of an external magnetic field and find an upper limit for the background interspecies two-body inelastic decay coefficient of $K^{\prime}_2 < 3\times 10^{-12}$cm$^3/$s for the ${^3P_2}$ $m_J=-1$ substate. We calculate the dynamic polarizabilities of the Yb(${^3P_2}$) magnetic substates for a range of wavelengths, and find good agreement with our measurements at $1064\,$nm. Our calculations also allow the identification of magic frequencies where Yb ground and metastable states are identically trapped and the determination of the interspecies van der Waals coefficients.

\end{abstract}

\maketitle

Ultracold elemental mixtures provide unique opportunities to study few- and many-body physics with mass-mismatched atomic partners \cite{kohs12} and diatomic polar molecules \cite{ni10,yan13}. While the bulk of elemental mixture experiments have been performed using ground-state bi-alkali systems, the recent production of ground state mixtures of alkali and alkaline-earth-like atoms \cite{nemi09,ivan11,hara11,pasq13} further extend the experimental possibilities. These include powerful quantum simulation and information protocols \cite{mich06} and tests of fundamental symmetries \cite{huds11} with paramagnetic polar molecules. While tunable two-body interactions that are important for these advances have been proposed in such mixtures \cite{zuch10}, they have not yet been experimentally detected.

In this Letter we report the realization a new class of heteronuclear mixtures in which one atomic component is in an electronically excited state, using lithium ($^6$Li) and ytterbium ($^{174}$Yb) atoms. This establishes a highly mass-mismatched atomic mixture where tunable anisotropic interactions are expected to play a strong role \cite{gonz13}, laying a foundation for future studies of ultracold trapped paramagnetic polar molecules and Efimov trimers with very large mass imbalance \cite{braa06}. We measure inelastic interactions in the mixture and observe the relative suppression of interspecies inelastic processes. Our experimental methods also demonstrate new techniques of production and manipulation of spin components in the metastable $^{3}P_2$ state of Yb.

\begin{figure}[!tb]
\includegraphics[angle = 0, width = 0.5 \textwidth] {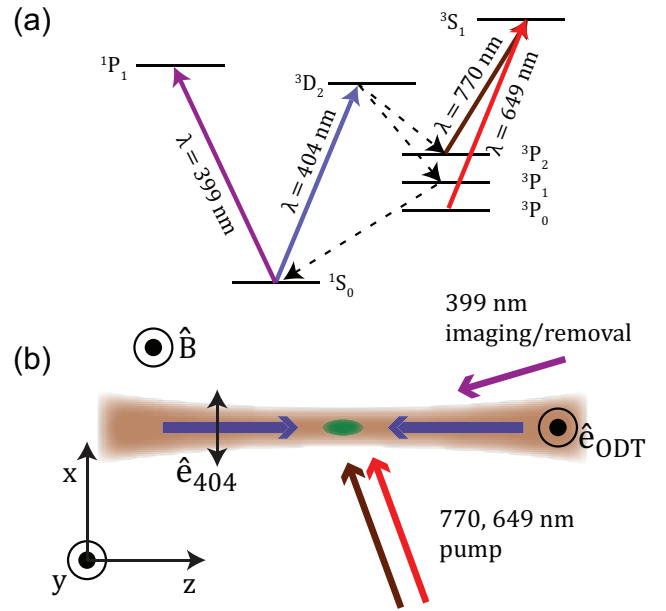} \vspace{-1.2cm} \caption{(Color online). Yb* preparation and detection scheme. (a) Low-lying energy levels of ytterbium with solid (dashed) lines indicating the relevant laser excitation (spontaneous decay) processes. The natural linewidths of the short-lived $\{^{1}P_1,^{3}P_1,^{3}D_2,^{3}S_1\}$ excited states are $2\pi\times\{28,0.18,0.35,12\}\,$MHz respectively. (b) Top-down view of experimental setup. Two counter-propagating $404\,$nm excitation beams (blue arrows) are overlapped with the ODT (brown). Laser beams at 649 and $770\,$nm (red, brown arrows) pump atoms  back to the ground state prior to absorption imaging using the $399\,$nm beam (purple arrow).}
\label{fig:fig1}
\end{figure}

The study and control of anisotropic interactions is an increasingly important topic in ultracold atomic systems. In addition to their impact on many-body physics \cite{mich06,laha09,yan13}, anisotropic two-body interactions are proving to be of great interest for generating magnetically tunable interactions, as has been calculated theoretically \cite{petr12} and observed experimentally in a mixture of ground and excited state Yb atoms \cite{kato13}. The latter result applied in the context of the Li+Yb combination points to an alternative route towards tunable interactions, where the ground state Feshbach resonances are predicted to be extremely narrow \cite{brue12} and experimentally difficult to access.

An important component of the work reported here is the successful trapping of Yb atoms in the $^{3}P_2$ state (Yb*) in a 1064nm optical dipole trap (ODT), where Li atoms can be co-trapped. Our scheme for preparation and detection of Yb* is similar to an earlier one that was used to populate a spin mixture of Yb* in a $532\,$nm ODT \cite{yama08} but is modified to produce pure spin states in a $1064\,$nm trap at an arbitrary external magnetic field. We confine ground state $^{174}$Yb atoms in a horizontally-oriented single-beam ODT and evaporatively cool them to the microkelvin regime. We then produce the Yb* state by optical pumping (see Fig.\ref{fig:fig1}) using the $^{1}S_0 \rightarrow {^3D}_{2}$ electric quadrupole transition at 404$\,$nm \cite{bowe99}. With this setup \cite{supp}, we can achieve a $^{1}S_0 \rightarrow {^3P}_{2}$ pumping rate of up to $50\,$Hz per atom. Remaining ground state atoms are removed with $399\,$nm light. Yb* atoms are detected by transferring them back to the ground $^{1}S_0$ state using light at $770/649\,$nm immediately prior to absorption imaging on the $399\,$nm transition. Further details can be found in the supplemental material.

\begin{figure}
\includegraphics[angle = 0, width = 0.5 \textwidth] {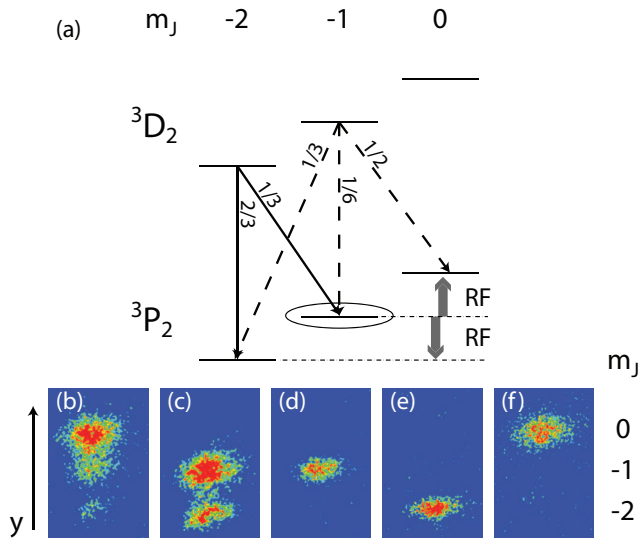}
\vspace{-0.5cm} \caption{(Color online). Yb* state preparation. (a) A particular magnetic substate of $^{3}D_{2}$ can be resolved via Zeeman splitting and selectively excited with $404\,$nm light. This state will subsequently spontaneously decay to substates in the ${^3P}_{2}$ manifold according to the indicated branching ratios. Radio frequency transitions within the $^{3}P_{2}$ manifold are spectroscopically resolved due to the state-dependent Stark shifts in the $1064\,$nm ODT. (b)-(f) Absorption images of different $^{3}P_{2}$ spin compositions after Stern-Gerlach separation. The images correspond to: (b) transfer via the $^{3}D_{2}$ $m_J=-1$ substate; (c) transfer via the $^{3}D_{2}$ $m_J=-2$ substate; (d) transfer via $^{3}D_{2}$ $m_J=-2$ substate with an applied in-trap gradient to obtain a pure sample of $^{3}P_{2}$ $m_J=-1$ atoms; (e) and (f) are the same as (d), followed by RF $\pi$-pulse transfer to $m_J=-2$ and $m_J=0$, respectively, of the $^{3}P_{2}$ manifold. The RF pulses in (e,f) are $300\,\mu$s long with frequencies close to the Zeeman splitting of $26\,$MHz but separated by $430\,$kHz due to the state-dependent Stark shift.}
\label{fig:fig2}
\end{figure}

Our method of preparing pure spin states of Yb* takes advantage of dipole selection rules and differential Stark shifts between spin states. By using an external magnetic field, we spectrally resolve magnetic substates within the ${^3D}_{2}$ manifold, from which the atoms decay into ${^3P}_{2}$ (see Fig.\ref{fig:fig2}(a)). The resultant population in a particular spin state of $^3P_2$ is determined by a combination of the branching ratio for spontaneous emission into that state and potential atom loss due to a trap depth reduction from the state dependent polarizability. We determine the spin composition of our trapped Yb* samples using the Stern-Gerlach technique to spatially separate the states during expansion from the trap [see Fig. \ref{fig:fig2}(b-f)]. The initial composition of atoms trapped in ${^3P}_{2}$ from an excitation to the $^{3}D_{2}$ $m_J=-1\,(-2)$ state is shown in Fig.\ref{fig:fig2}(b(c)). For our trap, the $^{3}P_{2}$ $m_J=-2$ is very weakly trapped compared to the $^{3}P_{2}$ $m_J=-1$ state, leading to a substantial, ODT beam power-dependent loss of $m_J=-2$ atoms due to gravitational spilling. By reducing ODT power and applying a vertical magnetic gradient field during the $404\,$nm exposure, we can make this state un-trapped. We thus obtain a pure sample of $m_J=-1$ Yb* atoms (Fig.\ref{fig:fig2}(d)). By applying a transverse radiofrequency (RF) magnetic field, we can drive transitions to other $m_J$ states of Yb*. Taking advantage of the spectroscopic resolution created by the differential Stark shift of neighboring states, we obtain pure samples of either $m_J=-2$ or 0 using RF $\pi$-pulses (Fig.\ref{fig:fig2}(e,f)) \cite{uenote}.

\begin{figure}
\includegraphics[angle = 0, width = 0.5 \textwidth] {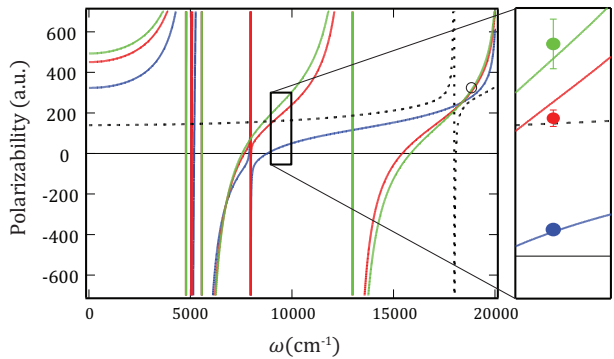}
\vspace{-1.2cm} \caption{(Color online). Calculated dynamic polarizability of Yb* as a function of frequency of light polarized parallel to the quantization axis. Green, red and blue solid lines correspond to the $|m_J|\,=\,$0, 1, and 2 substates respectively and the dashed line to the ground state. The open circle indicates measurements of the ac polarizability at $532\,$nm \cite{yama10}. Our polarizability ratio measurements at $1064\,$nm, scaled to the calculated $^1S_0$ value, are shown as solid circles in inset.}
\label{fig:fig3}
\end{figure}

The relevant property for controlling a particular atomic state in an optical trap is its dynamic polarizability $\alpha(\omega,\vec{\epsilon})$,
a function of radiation frequency $\omega$ and polarization $\vec{\epsilon}$. The polarizability of each of the ${^1S}_0$ ground and ${^3P}_2$ metastable states of Yb is determined by the dipole couplings to all other atomic states \cite{Koto06}. We calculate the polarizability from a combination of experimental transition frequencies and oscillator strengths between atomic levels available from the literature \cite{nist11} and additional theoretical calculations of these characteristics for other transitions using a relativistic multiconfiguration Dirac-Fock-Sturm method, described in \cite{Koto07}. Figure \ref{fig:fig3} shows the resulting polarizability as a function of laser frequency of light linearly polarized parallel to the quantization axis for ground state Yb and the five magnetic sublevels $m_J$ = 0, $\pm$ 1, and $\pm$ 2 of Yb*. The polarizability is singular at atomic transition energies and strongly depends on the absolute value of $m_J$. In fact, it has opposite signs for different $m_J$ over significant ranges of frequencies. The intersections of ground and excited state curves in Fig.$\,$\ref{fig:fig3} indicate magic wavelengths for ultra-narrow optical transitions. We find good agreement with earlier measurements at $532\,$nm \cite{yama10}. The transition frequencies and oscillator strengths used in calculating the dynamic polarizability also enable us to construct both relativistic and non-relativistic van der Waals $C_6$ coefficients for the Yb*+Yb* and Yb*+Li collision systems.  A more detailed description of this procedure is given in the supplemental material. From these we determine the $p$-wave threshold for two-body collisions in the above systems to be 24$\,\mu$K and 2.1$\,$mK, respectively.

We experimentally determine polarizabilities at 1064 nm to compare to our calculated values. We measure the trap frequencies of the $m_J=0,\,-1$ and $-2$ substates and compare to an identical measurement for the ground $^{1}S_0$ state. The trap frequencies were measured by observing the oscillation of cloud size (breathing mode) following a diabatic increase in trap depth. These values were also verified by using the parametric excitation technique. The experimentally obtained polarizability ratios are $\alpha_{-1}/\alpha_{g}=1.04(6)$, $\alpha_{0}/\alpha_{g}=1.6(2)$, and $\alpha_{-2}/\alpha_{g}=0.20(2)$. These agree well with our theoretically calculated ratios (see inset of Fig.\ref{fig:fig3}).

With the facility to produce, manipulate, and detect Yb* added to our Li-Yb apparatus \cite{ivan11,hans13}, we investigate the mixture of Yb* and Li. Here, we focus on a mixture of the lowest hyperfine state of $^6$Li (denoted $|1\rangle$) and the $m_J=-1$ state of Yb*. Starting with an optically trapped and cooled mixture of $^{174}$Yb (${^1}S_0$) and $^{6}$Li ($|1\rangle$), we change the magnetic field to a desired value and prepare the $m_J=-1$ state of Yb* as discussed above \cite{supp}. We subsequently compress the trap in order to suppress atom loss due to evaporation and to improve interspecies spatial overlap against differential gravitational sag. Further details can be found in the supplemental material.

\begin{figure}
\includegraphics[angle = 0, width = 0.5 \textwidth] {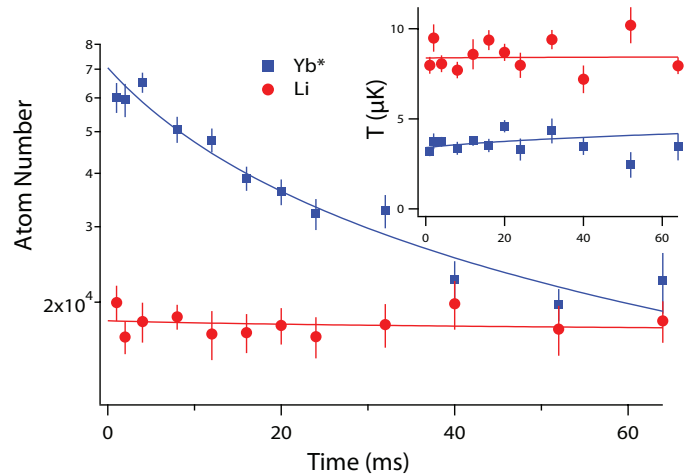}
\vspace{-0.5cm} \caption{(Color online). Number and temperature (inset) evolution of Yb* and Li in the ODT at an external magnetic field of 12$\,$G. Lines show fits based on a model including two-body inelastic effects (see text).}
\label{fig:fig4}
\end{figure}

The number and temperature evolution of an Yb*+Li mixture prepared in this way at an external magnetic field of $12\,$G is shown in Fig.\ref{fig:fig4}. The initial temperature difference between the species is due to thermal decoupling at the lowest trap depths \cite{ivan11,supp} and is a useful starting point to monitor elastic interspecies interactions. One-body effects from Yb* spontaneous decay, from collisions with background atoms, and from off-resonant scattering of ODT photons are negligible on the timescale of the experiment. Since our lithium component is a single state fermion deep in the $s$-wave regime \cite{yan96}, the Li-Li interactions are negligible for our parameters. From our calculated values for the $p-$wave threshold for two-body collisions in the Yb*+Yb* and Yb*+Li systems, we infer that all two-body interactions are $s$-wave dominated. For the starting peak density of $n_{\rm Yb^*(Li)} = 5.3 (1.2)\,\times\,10^{12}\,$cm$^{-3}$ and large evaporation parameter $\eta_{\rm Yb^*(Li)} = {U_{\rm Yb^*(Li)}}/{k_B T_{\rm Yb^*(Li)}} = 24(20)$, it is reasonable to expect that all number losses result from two-body inelastic processes. Here $U$ and $T$ are the trap depths and temperatures of the two species, respectively. We observe that the system retains its initial temperature disparity throughout the timescale of the experiment. This suggests that interspecies $s-$wave elastic collisions play a negligible part in the system dynamics.

A striking feature of Fig.$\,$\ref{fig:fig4} is the stability of the Li number on the timescale of Yb* decay, suggesting a dominance of inelastic effects from Yb*+Yb* collisions over Li+Yb* collisions. We model the atomic densities with the coupled differential equations
\begin{eqnarray}
\dot{n}_{\rm Yb^*} & = & -K^{\prime}_2 n_{\rm Li}n_{\rm Yb^*}-K_2 n_{\rm Yb^*}^{2}\\
\dot{n}_{\rm Li} & = & -K^{\prime}_2 n_{\rm Li}n_{\rm Yb^*}
\end{eqnarray}
where $n_{\rm Yb^*(Li)}$ and $K_2(K^{\prime}_2)$ are the densities of Yb*(Li) and the (volume-independent) two-body decay coefficients of Yb*-Yb*(Yb*-Li).
The temperature evolution is given by the heating from the density-dependence of the inelastic processes which favor atom loss from near the trap center \cite{webe03}. Best-fits with this model (solid lines in Fig \ref{fig:fig4}) yield $K_2 = 2.5\times 10^{-11}\,$cm$^3$/s, and $K^{\prime}_2$ consistent with zero \cite{k2note}. The estimated statistical error provides an upper bound of $K^{\prime}_2 < 3\times 10^{-12}$cm$^3/$s. By considering the elastic cross-section needed for interspecies thermalization on the experimental timescale, we can also place an upper bound of $300\,a_0$ on the magnitude of the interspecies $s-$wave scattering length. Here $a_0$ is the Bohr radius.

Our study of interspecies interactions can be extended to arbitrary values of the external magnetic field and also to different magnetic substates of Yb* using the methods described above. Repeating the above experiment at $94\,$G we observe a similarly long lifetime of Li atoms in the presence of Yb*. These low values of background interspecies inelastic rates bode well for future searches for Feshbach resonances between spin-polarized samples of Yb* and Li, where interspecies inelastic rates should be resonantly enhanced and could be observed by monitoring the Li population as a function of magnetic field. By working in an optical lattice and/or using a fermionic Yb isotope, the inelastic effects of Yb*+Yb* collisions may be suppressed, allowing for a more precise investigation of interspecies phenomena in the mixture. We note that complementary theoretical work has already been initiated \cite{gonz13,koto13}.

In conclusion, we produced spin polarized samples of ytterbium atoms in the ${^3P}_{2}$ (Yb*) state in a 1064 nm ODT. Our demonstrated method to manipulate the spin state of Yb* is extendable to production of arbitrary spin superpositions within the ${^3P}_{2}$ manifold and could be applicable towards quantum information schemes reliant on the long-range magnetic dipole-dipole interaction \cite{dere04}. We measured the dynamic polarizabilities of different spin substates and found good agreement with our calculated theoretical values. These calculations also identify magic wavelength points, of relevance for potential optical clock transitions between ground and long-lived metastable (${^3P}_2$) states of Yb \cite{hall89,lemk09}. We co-trapped lithium with metastable ytterbium and investigated the interaction properties of the mixture at large magnetic fields. We found a dominance of intraspecies inelastic effects over interspecies ones. Possible applications of this mixture include investigations of universal few-body physics \cite{braa06}, and the synthesis of ultracold paramagnetic polar molecules for quantum simulation \cite{mich06} and tests of fundamental symmetries \cite{huds11}.

We thank J.M. Hutson, P.S. Julienne and Y. Takahashi for useful discussions, and A.O. Jamison for valuable technical contributions and discussions. We gratefully acknowledge funding from NSF grants PHY-1308573, PHY-0847776, PHY-1306647, AFOSR grants FA9550-11-1-0243, FA9550-12-10051, and ARO MURI grant W911NF-12-1-0476.

{\it Note added:} After submission of our paper, related work was reported in which the polarizabilities of Yb* substates were determined at 1070nm using optical spectroscopy \cite{hara13}.


\begin{thebibliography}{33}
\bibitem[{uen(2013)}]{uenote}
\bibinfo{journal}{Another method to achieve pure spin states in Yb* has been
  shown earlier \cite{ueta12}, but relies on the availablity of an ultra-narrow
  laser}.
\bibitem[{k2n(2013)}]{k2note}
\bibinfo{journal}{By fitting the Yb* number according to $\dot{n}_{\rm Yb^*} =
  -K_2 n_{\rm Yb^*}^{2}$ at the average temperature of $3.6\,\mu$K, we obtain
  the value of $K_2 = 2.6 \times10^{-11}\,$cm$^3/$s.}
\end{thebibliography}
\end{document}


{\bf Supplemental Material}

\vspace{5mm}

{\bf Metastable State Preparation and Detection}

The 404$\,$nm laser frequency is stabilized using a Fabry-Perot cavity which is in turn stabilized to the $^{1}S_0 \rightarrow {^1P}_{1}$ laser at 399$\,$nm, forming a transfer-lock setup, similar to the one described in \cite{ueta08}. The ${^3D}_{2}$ state has a linewidth of $2\pi\times350$ kHz, with branching fractions of 0.88 and 0.12 to ${^3P}_{1}$, and ${^3P}_{2}$, respectively. Thus each ground state atom is pumped to the ${^3P}_{2}$ state after an average of $1/.12\approx8$ excitations. To optimize the pumping efficiency, the available $1\,$mW of $404\,$nm light is tightly focused to a waist of about $25\,\mu$m at the atomic sample and aligned along the axial direction of the ODT. Two counter-propagating laser beams are used to suppress axial dipole oscillations of the Yb* cloud due to photon recoil. The initial transfer rate is $50\,$Hz per atom, but the overall transfer efficiency is limited by inelastic loss. For all the experiments discussed, we remove any remaining $^{1}S_{0}$ atoms after transfer with a $3\,$ms light pulse resonant with the ${^1S}_{0}\rightarrow{^1P}_{1}$ transition.

The transfer of Yb* to the ground state for detection is achieved by applying a $200\,\mu$s pulse of $770\,$nm light resonant with the $^{3}P_{2} \rightarrow {^3S}_{1}$ transition. The ${^3S}_{1}$ state has a $13\,$ns lifetime and decays via the ${^3P}_{1}$ state to ${^1S}_0$. Atoms which decay to the long-lived ${^3P}_{0}$ state can be brought back into the transfer cycle with a simultaneously applied $649\,$nm pulse resonant with the $^{3}P_{0} \rightarrow {^3S}_{1}$ transition. We typically used only the $770\,$nm resonant light, resulting in some atoms getting stuck in ${^3P}_0$. The fraction of such atoms was calibrated by comparing detection with and without the $649\,$nm beam and found to be $0.24(1)$. This is in reasonable agreement with the $0.26$ value for the branching fraction obtained from the $^{3}S_{1} \rightarrow {^3P}$ reduced matrix elements calculated in \cite{pors99}.

\begin{figure}
\includegraphics[angle = 0, width = 0.5 \textwidth] {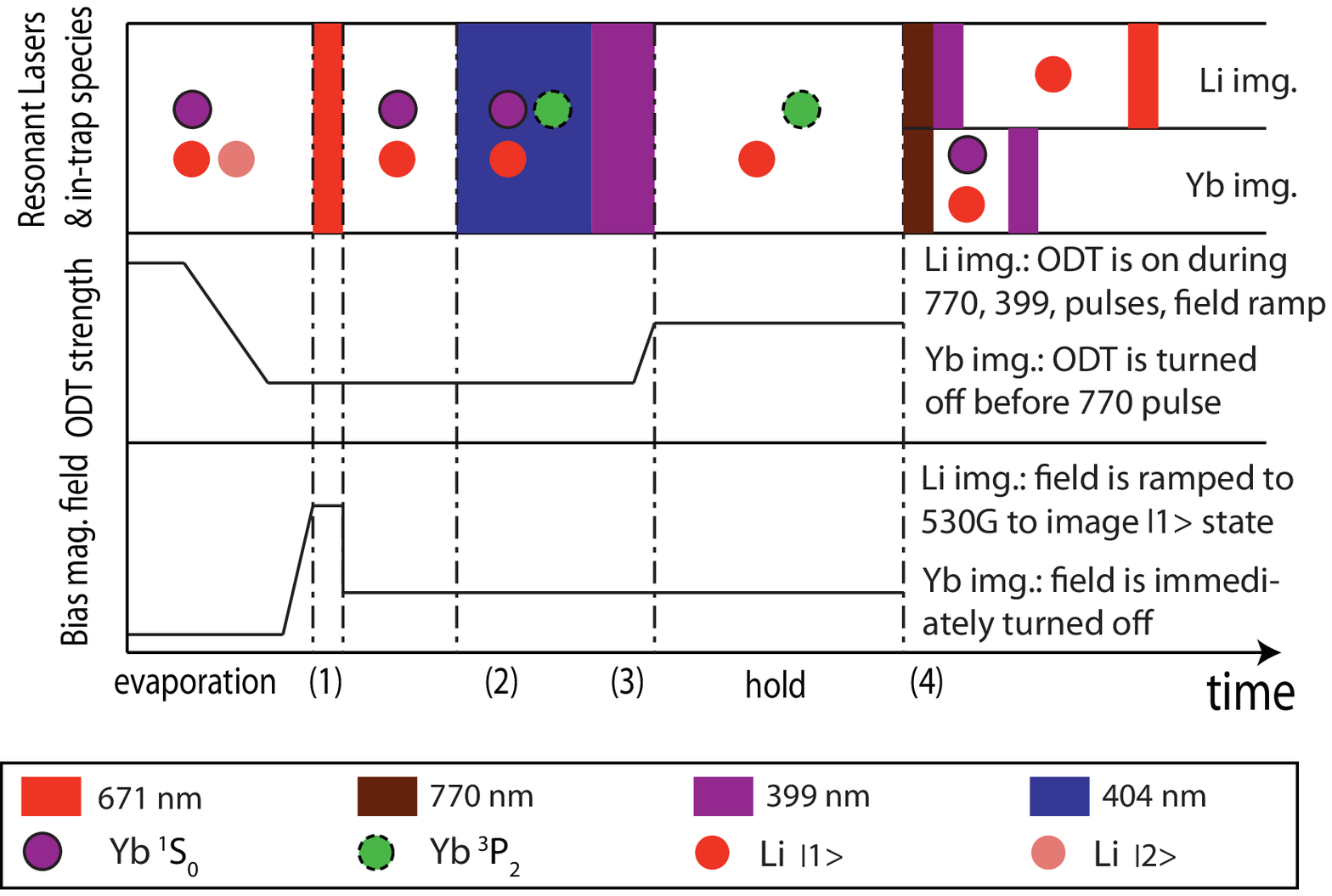}
\caption{(Color online). The timeline of the dual-species experiment. After evaporation, we remove the $\left|2\right\rangle $ state of lithium over $10\mu$s with resonant light at $671\,$nm (1), then go to the desired magnetic field, and excite with $404\,$nm over $20\,$ms (2). Following the end of that excitation, we remove the remaining ${^1S}_{0}$ atoms in $3\,$ms with a resonant $399\,$nm pulse (3). After a variable hold time, we transfer the metastable atoms back to the ground state via ${^3S}_{1}$ and then image (4), as described.}
\label{fig:figs2}
\end{figure}

\vspace{5mm}

{\bf Mixture Preparation}

For preparation of the Li+Yb* mixture, we begin with an initial ODT depth of $U_{\rm Yb}=560\,\mu$K. We evaporatively cool Yb by lowering the trap depth to $15\,\mu$K in $3\,$s. The Yb cloud sympathetically cools an equal mixture of the two lowest hyperfine ground states of Li, which experience a $2.2$ times greater optical potential at $1064\,$nm. We spin-purify the Li component by energetically resolving the two states (denoted $|1\rangle$ and $|2\rangle$) at a high magnetic field ($480\,$G) and removing the upper state ($|2\rangle$) from the trap with resonant light. After preparing Yb*, we remove any remaining $^{1}S_{0}$ atoms with a $3\,$ms light pulse resonant with the ${^1S}_{0}\rightarrow{^1P}_{1}$ transition, the last $1\,$ms of which is concurrent with a re-compression step where the trap depth is ramped-up to $U_{\rm Yb}=85\,\mu$K, corresponding to a mean trap frequency of $\bar{\omega}_{\rm Yb^*(Li)}=2\pi \times 220 (1700)\,$Hz. The re-compression step is done in order to suppress atom loss due to evaporation and to improve interspecies spatial overlap against differential gravitational sag. In the absence of the Yb* preparation step, the Yb-Li mixture contains $N_{\rm Yb(Li)}=2.3(0.5)\times10^5$ atoms at temperature $T_{\rm Yb(Li)}=4 (8)\,\mu$K. The temperature difference between the species is due to thermal decoupling at the lowest trap depths \cite{ivan11}. The entire experimental sequence for the mixture studies is diagrammed in Fig. \ref{fig:figs2}.

It was important to consider the residual motion of Yb* in the ODT. The optical pumping step deposits several photon recoils of energy into the sample along the ODT axis while the differential Stark shift results in a small but instantaneous change in the trap strength. Additionally, the timescale of the re-compression step (limited by the inelastic lifetime of Yb*) is fast compared to the axial frequency of Yb* in the ODT. These often led to axial dipole oscillations of the Yb* atoms in the trap in a single $404\,$nm beam setup. We mitigated this effect by using two balanced counterpropagating $404\,$nm beams. There were still some residual damped breathing oscillations, the amplitude of which varied between different datasets. Such differences were most likely due to differences in the amount of untransferred ground state Yb atoms that can collisionally damp out these excitations, before the $399\,$nm removal pulse. In order to avoid this complication, we chose to analyze a data set which did not exhibit this oscillatory behavior.

\vspace{5mm}

{\bf Theoretical Supplement}

The long-range van-der-Waals dispersion interaction is fully determined by the atomic dynamic polarizability of Li($^2S_{1/2}$) and Yb($^3P_2$) as a function of imaginary frequency \cite{Ston96}. The polarizability of Li is well described from the experimental data given in \cite{nist11}. In the main paper, we only give details of our determination of the polarizability of the metastable Yb states.

\begin{table}
\caption{Non-relativistic and relativistic $C_6$ dispersion coefficients
in units of $E_ha_0^6$ and magnetic dipole-dipole interaction coefficients
$C_3$ in units of $E_ha_0^3$ between a ground-state Li and a metastable
$^3P_2$ Yb atom. The non-relativistic coefficients are labeled by
$^{2S+1}\Lambda$, the relativistic by $n(\Omega)$. Here, $E_h=4.35974
\times 10^{-18}$ J is the Hartree and $a_0=0.0529177$ nm is the Bohr
radius.}

\begin{tabular}{cc|ccc}
\hline
$\Lambda$&~~~~ $C_6$~~~~&$n(\Omega)$ &~~~~$C_6$~~~~ &~~~~  $C_3 (10^{-5}) $~~~~~    \\
       \hline
 $^{2,4}\Sigma$  & 3279.87   & 1(1/2)   &  2987.57  & 3.2902   \\
 $^{2,4}\Pi$     & 2402.98   & 2(1/2)   &  2841.42  & -7.2886  \\
                 &           & 1(3/2)   &  2841.42  & 5.2091   \\
                 &           & 2(3/2)   &  2402.98  & -9.2075  \\
                 &           & 1(5/2)   &  2402.97  & 7.9969
\end{tabular}
\label{coef}
\end{table}

The transition frequencies and oscillator strengths used in obtaining the dynamic polarizability enable us to construct both relativistic and non-relativistic van der Waals $C_6$ coefficients. They are listed in Table~\ref{coef}.  The relativistic coefficients, which are directly obtained using dynamic polarizability at imaginary frequencies \cite{Koto10}, describe the long-range interaction potentials with symmetry $n(\Omega)$, where $\Omega$ is the projection
of the total electron angular momentum along the internuclear axis and $n$ indexes states for the same $\Omega$.  For $\Omega$ = 1/2, 3/2, and 5/2 there are 2, 2, and 1  adiabatic relativistic Born-Oppenheimer (BO) potentials dissociating to the Li$(^2S_{1/2})$ + Yb$(^3P_2)$ limit, respectively. The non-relativistic $C_6$ coefficients with symmetry $^{2S+1}\Lambda$ are found from the relativistic coefficients assuming an $\vec l_a\cdot \vec s_a$ coupling between the electron orbital angular momentum $\vec l_a$ and spin $\vec s_a$ for both Li ($a=Li$) and meta-stable Yb ($a=Yb$). Our two non-relativistic coefficients differ significantly from the coefficients used in \cite{gonz13}.